\documentclass[12pt]{article}
\usepackage{times}
\usepackage{graphicx}
\usepackage{multirow}
\usepackage{color}
\usepackage{bm}
\usepackage{soul} 
\usepackage[separate-uncertainty=true,number-mode=math,unit-mode=text]{siunitx}
\usepackage{tabularx}
\usepackage{physics}

\title{Evolution of dipole-dipole dynamics in cold ammonia collisions}

\author
{Yao Chang$^{1}$, Andr\'{e} J.A. van Roij$^{1}$, Stach Kuijpers$^{1}$, Sven Herbers,\\Etienne F. Walraven$^{1}$, Tijs Karman$^{1\ast\ast}$, Sebastiaan Y.T. van de Meerakker$^1$$^{\ast\ast}$ \\
\normalsize{$^1$Radboud University, Institute for Molecules and Materials}\\
\normalsize{Heijendaalseweg 135, 6525 AJ Nijmegen, the Netherlands}\\
\normalsize{$^{\ast\ast}$To whom correspondence should be addressed;}\\
\normalsize{E-mail: basvdm@science.ru.nl, t.karman@science.ru.nl}}

\begin{document}
\date{\today}

\maketitle

\begin{abstract} 
Cold polar molecules offer fascinating prospects for ultracold chemistry and quantum physics, including new platforms for quantum simulation or computation. However, their inherent collision properties remain largely unknown. It has proven extremely hard to experimentally probe collisions between two dipolar molecules at sufficiently low energies and high precision, as it appears fundamentally impossible to merge two beams of molecules with significant dipole moments. Here we report measurements of state-to-state cross sections for collisions between ammonia isotopologues at energies between 0.3 and 100 cm$^{-1}$ using a novel beam merger. We experimentally observed a local maximum in the cross sections that indicates the effective dipole moments can switch off at low collision energies. Scattering calculations reproduced this maximum in good agreement and explained the observed scaling with the parity splitting energies in the molecular energy level structures. Measurements of the correlated energy transfer in both collision partners yielded direct evidence of the suppression of the dipole-dipole interaction at low energies. Our results demonstrate how collisions between an important class of polar molecules evolve from the high temperature limit towards the ultracold regime in a counterintuitive way, have major consequences for the feasibility of future experiments and the interpretation of previous work, and offer distinctive opportunities to control cold molecular collisions with external fields.
\end{abstract}%

\newpage
The topic of low-energy collisions between polar molecules is among one of the most exciting currently studied in molecular physics and physical chemistry \cite{Langen:NatPhys20:702,Karman:NatPhys20:722,Bohn:Science357:1002,
Heazlewood:NatRevChem5:125,Bell:MolPhys107:99,Dulieu:Book2018,Balakrishnan:JCP145:150901}. The strong and long-range dipole-dipole interaction offers distinctive prospects to control intermolecular dynamics with external electric or magnetic fields \cite{Ospelkaus:Science327:853,Ni:Nature464:1324,Li:NatPhys17:1144,Li:Nature614:70,Krems:IntRevPhysChem24:99,Koller:PRL:203401,Tscherbul:PRL115:023201,Krems:PCCP10:4079}, whereas trapped samples of ultracold polar molecules promise the realization of new phases of matter \cite{Langen:PRL134:053001} and new paradigms for quantum simulation or computation \cite{Cornish:NatPhys20:730,DeMille:PRL88:067901,Gregory:NatPhys17:1149,Ruttley:Nature637:827}. Yet, even after two decades of method development that enables novel collision experiments \cite{Sawyer:PCCP13:19059,Kirste:Sience338:1060,Parazzoli:PRL106:193201,Tang:Science379:1031} as well as the preparation of trapped cold \cite{Weinstein:Nat395:148,Meerakker:CR112:4828,Segev:Nature572:189,Stuhl:NATURE492:396}, ultracold \cite{Softley:PRSA479:20220806,Park:NatPhys19:1567,McCarron:PRL121:013202,Ni:Science322:231,Barry:Nature512:286} or even quantum degenerate samples of polar molecules \cite{DeMarco:Science363:853,Bigagli:Nature631:289,Valtolina:Nature588:239}, their inherent collisional properties remain largely unkown. Theoretical \emph{ab initio} methods generally lack quantitative predictive power at low energies, and it is not clear how classical theories typically used at high energies connect to quantum mechanical descriptions needed at low energies. Several theoretical models predict novel collision dynamics induced by the dipole-dipole interaction \cite{Tang:Science379:1031,Avdeenkov:PRL90:043006,Bohn:NJP11:055039,Wang:PRR6:L022033,Sykes:PRA91:013625}, but precise experimental collision data to validate these predictions remains scarce across all energy domains. Our lack of understanding of the inherent collision properties of dipole-dipole governed systems is incommensurate with its large importance to the fields of cold and ultracold molecules, and urgently needs solving.  

To appreciate the novelty of low-energy dipolar collisions, one must first understand the nature of the molecular electric dipole moment itself. In classical mechanics and electrostatics, the occurrence of a dipole moment $d$ in a molecule originates from an asymmetric charge distribution within the molecule. The resulting permanent dipole moment is key to understand reactivity in chemistry, and  physical properties of bulk systems such as solubility, and boiling or melting points. In condensed water (H$_2$O), one of the most well-known polar molecules, the dipole moment underlies hydrogen bonding that plays a fundamental role in chemistry. 

Within this classical picture, collisions between two isolated dipolar molecules can be approximated by the Langevin capture model \cite{Levine:reaction-dynamics}. One can analytically show that the collision cross section only depends on the dipole moments involved, and scales with the collision energy $E_{coll}$ as $E_{coll}^{-2/3}$, i.e., the cross section increases as the collision energy or temperature decreases. This model has been remarkably effective in describing strongly interacting systems, and is used in the community to predict scattering cross sections for collisions between cold polar molecules throughout \cite{Soley:CR125:6609}.     

Yet, in a quantum-mechanical description of the molecule, the nature of the dipole moment is more subtle. Here, the molecule is described by a wavefunction that satisfies Schr\"{o}dinger’s equation. This wavefunction has a property, called parity $p$, that has no classical analogue and that can either be even ($+$) or odd ($-$). The parity describes how the wavefunction transforms upon inversion of all coordinates, and therefore embeds information on the symmetries of the molecule itself. In contrast to the classical picture, this implies that any isolated molecule in a given non-degenerate quantum state with definite parity must have zero dipole moment in the laboratory frame \cite{Note:5}. A dipole moment can then only arise if the wavefunctions of two states with opposite parity are mixed, as occurs for instance in the presence of an external electric field or by interactions with the environment. For most molecules, mixing between adjacent rotational levels with opposite parity has to be achieved \cite{Ospelkaus:Science327:853,Ni:Nature464:1324,Vilas:arXiv:2025}. Indeed, molecules are often called “polar” if significant mixing can already be achieved in moderate electric field strengths. 

Although the difference between the classical and quantum description of a dipole moment may seem subtle and academic, it can have huge consequences for the scattering behaviour of polar molecules. Whereas in the classical Langevin picture one predicts the cross sections to always increase as the temperature is reduced, the quantum nature of the dipole moment can lead to intricate collision dynamics with a counterintuitive behaviour of the cross sections that defies common wisdom. This was recently predicted to occur in a special class of molecules where each rotational level is split into two near-degenerate components with opposite parity, as is typically found in chemically relevant molecules such as open-shell radicals like OH and NO, asymmetric tops like formaldehyde (H$_2$CO), molecules with double-well potentials like ammonia (NH$_3$), or molecules with a parity doublet originating from coupling with vibrational bending modes like CaOH \cite{Vilas:nature606:70,Vilas:arXiv:2025,Robichaud:Nature2026}. The parity doublet structure gives rise to a previously unknown scattering phenomenon, resulting in a characteristic local maximum (LM) in the scattering cross section \cite{Tang:Science379:1031}. At high collision energies, the molecules come sufficiently close together that the energy associated with the dipole-dipole interaction, which by its nature couples states with opposite parity, exceeds the parity doublet splitting energy $\Delta_A$ and $\Delta_B$ of the two colliders. This causes a mutual polarization of both molecules as they approach each other, i.e., the parity wavefunctions become mixed even in the absence of any external electric field. This self-polarization effectively switches on the dipole moments of the molecules, and the Langevin capture cross section is retrieved. As the energy is reduced, however, molecules do not come sufficiently close anymore to sustain this mutual polarization. This effectively switches off the dipole moments, leading to a cross-over regime where the Langevin capture model breaks down and the cross sections diminish by up to orders of magnitude. As the energy is reduced even further towards the ultracold regime, the cross section increases again obeying the Wigner threshold law.  

The occurrence of a LM in the cross section was predicted to be ubiquitous for any combination of molecules within this class, and has monumental impact on the feasibility of future experiments as well as the interpretation of previously obtained results alike \cite{Wu:Science358:645}. The much lower than anticipated (inelastic) cross sections may impact efforts to reduce the temperature of trapped ensembles of polar molecules via evaporative cooling that rely on the ratio between inelastic and elastic cross sections \cite{Stuhl:NATURE492:396}. Furthermore, in efforts to study the inherent properties of cold collisions in crossed beam experiments, for instance, the breakdown of the Langevin cross section incline towards lower energies quickly impedes state-to-state cross section measurements. 

To date, the LM and its underlying collision mechanism remain a theoretical prediction. This raises pressing questions about the true nature of an important class of cold dipolar collisions: Does the LM indeed exist, or are there other effects neglected by theory models that overshadow it? Is the existence of the LM indeed universal and ubiquitous for these kind of systems? What happens at energies below the LM where the dipole moments have effectively switched off, and which mechanisms dominate the collision then? And how does the LM change if external fields are added that modify both parity states independent of the collision energy? 

Answering these questions experimentally is extremely difficult, however. The collision energy at which the LM occurs was predicted to scale with $\Delta =\Delta_A + \Delta_B$.  As $\Delta \ll$ 1~cm$^{-1}$ \cite{Note:4} for most molecules, the LM occurs at very low collision energies that thus far have not been accessible experimentally. Recently, state-to-state inelastic collision cross sections were measured for the NO ($d=0.16$ D, $\Delta_{\textrm{NO}}=0.01$~cm$^{-1}$) - ND$_3$ ($d=1.5$ D, $\Delta_{\textrm{ND}_3} = 0.05$ cm$^{-1}$) system in a merged beam configuration by bending the ND$_3$ beam into the NO beam's path, but the minimum collision energy of 0.15~cm$^{-1}$ achieved was still too large to probe the LM \cite{Tang:Science379:1031}. Reaching lower collision energies appears unfeasible using currently available methods, whereas change of system with larger $\Delta$ to move the LM to higher energies is also met by impeding experimental hurdles. One fundamental limitation is that the merged beam technique cannot be used for two molecules with a similar dipole moment and mass, as bending one beam will necessarily deflect the other in opposite direction preventing sufficient overlap for collisions to occur. This prevented the use of (combinations of) OH, NH$_3$, and H$_2$CO which would for their values for $d$ and $\Delta$ be the natural choice to probe the LM.  

Here, we present a joint experimental and theoretical study on inelastic collisions between isotopologues of ammonia. Using a combination of Stark deceleration, curved hexapole guiding, and a novel beam merger that allowed for low collision energies despite the fundamental limitation in merging two beams of polar molecules with similar dipole moments, we measured fully state-resolved integral and differential cross sections for the ND$_3$-ND$_3$ ($\Delta=0.1$ cm$^{-1}$), ND$_3$-NH$_3$ ($\Delta=0.85$ cm$^{-1}$) and NH$_3$-NH$_3$ ($\Delta=1.6$ cm$^{-1}$) systems at collision energies in the $0.3 - 100$ cm$^{-1}$ energy range. We observed the LM for each system, and found universality in the scaling of its peak position with $\Delta$. We demonstrated the ubiquity of the self-polarization mechanism by complementary measurements on NO-NH$_3$ ($\Delta=0.81$ cm$^{-1}$) that showed the LM also for this system. On the low-energy side of the LM, we observed a breakdown of the dipole-dipole interaction strength in favor of the dipole-quadrupole interaction that gained relative importance. We used high-resolution velocity map imaging (VMI) to separate the contribution of both interactions by recording the total recoil energy from de-excitation collisions imparted to the collision products. We observed that both collision partners underwent a parity flip when the dipole-dipole interaction dominated, whereas only one of the partners flipped for a dipole-quadrupole interaction. Good agreement was obtained with the cross sections derived from quantum coupled-channels scattering calculations, which revealed the delicate roles of $\Delta$, the dipole-dipole and dipole-quadrupole interaction potentials as well as the degeneracy of the energy level structure in both collision partners, that all can either support or discourage energy transfer. 
\\

\section*{Results and Discussion}

\begin{figure}[!htb]
	\centering
	\includegraphics[width=\linewidth]{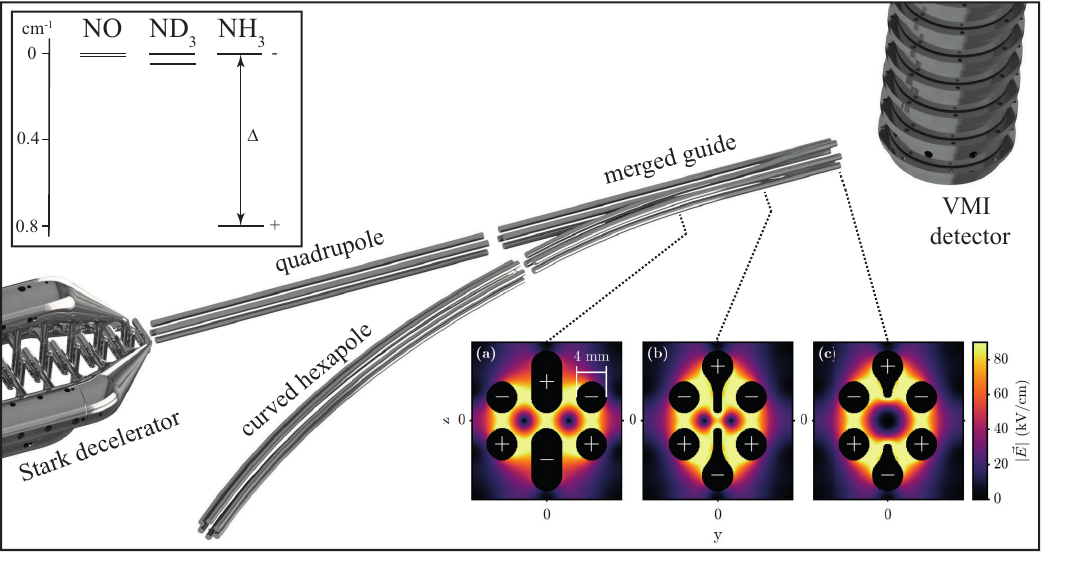}
	\caption{{\bf Schematic representation of the experimental setup.} A state-selected and velocity tunable beam of $^{14}$ND$_3$, $^{15}$ND$_3$, $^{15}$NH$_3$ or NO molecules is produced using a 2.6-meter-long Stark decelerator, and merged with a state-selected beam of $^{14}$ND$_3$ or $^{14}$NH$_3$ using a curved hexapole. Molecules from both beamlines are loaded into a merged guide in which two adjacent quadrupole traps gradually evolve into a single hexapole trap. Cross sections and electric field distribution of the merged guide at three different positions are shown in the inset (adapted from Ref. \cite{Kuijpers:RSI95:093201}). The molecules emerging from the guide predominantly collide with an effective beam crossing angle of 2$^{\circ}$ just downstream from the exit of the guide. Molecules are transported from the Stark decelerator to the entrance of the merged guide using a straight quadrupole segment. Only the last section of the decelerator is shown. Scattered molecules originating from the Stark decelerator beamline are detected state-selectively using velocity map imaging and recoil-free laser ionization schemes. The rotational ground state structure of NO, ND$_3$ and NH$_3$ are shown in the inset. Each level is split into two components of opposite parity with energy splitting $\Delta$, where $\Delta_{\textrm{NO}} \ll \Delta_{\textrm{ND}_3} \ll \Delta_{\textrm{NH}_3}$. }
\label{fig:setup}
\end{figure}

We used a crossed molecular beam apparatus that is schematically shown in Fig. \ref{fig:setup} and explained in more detail in the Supporting Materials (SM). Packets of ND$_3$ or NH$_3$ in the $X\, ^1A_1^{\prime}, v_2=0, j_k^p=1_1^-$ state (referred to hereafter as the $|-\rangle$ state) with a tunable velocity were produced by passing a beam of $\sim$ 5 \% ammonia seeded in argon, krypton or xenon through a 2.6~m long Stark decelerator. As a collision partner, beams of ND$_3$ or NH$_3$ seeded in xenon were passed through a curved hexapole. Both beams were then loaded into a merged guide, in which the electric fields generated by a geometry of 6 individual electrodes gradually evolved from two adjacent quadrupole traps into a single hexapole trap \cite{Kuijpers:RSI95:093201}. This guide was optimized to launch the two near-copropagating packets into a region with near-zero electric field where they overlapped with an effective crossing angle of 2$^{\circ}$, yielding a minimum collision energy of $\sim 0.3$~cm$^{-1}$. This near-merging approach yielded maximum spatial overlap between the beams at minimum collision energy, and limited undesired partial overlap in the last section of the guide where the large inhomogeneous electric fields could strongly affect the collision cross sections and interpretation of the results. Molecules that scattered into the $1_1^+$ lower inversion doublet component of the $j_k=1_1$ level, referred to hereafter as the $|+\rangle$ state, were state-selectively detected and velocity mapped on a two dimensional detector \cite{Plomp:MP119:e1814437}. We implemented a two-color laser ionization scheme involving a vacuum ultraviolet (VUV) laser developed recently to eliminate blurring of the images due to ion recoil \cite{Kuijpers:JPCA128:10993}. We used isotope labeling of the N atom to spectroscopically separate signals from the packets emerging from the straight and curved arms of the experiment.\\

\begin{figure}[!htb]
	\centering
	\includegraphics[width=0.8\linewidth]{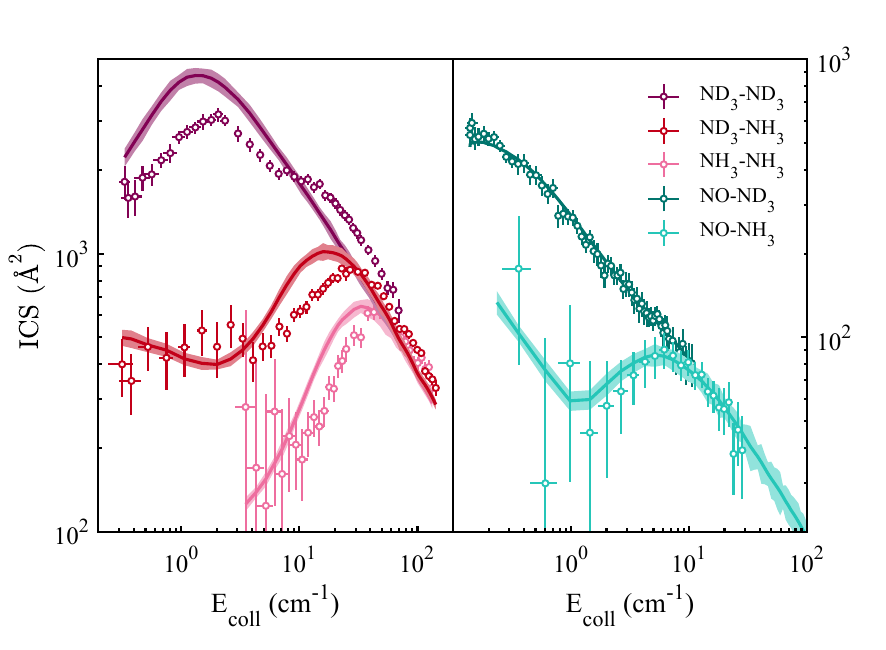}
	\caption{{\bf Collision energy dependence of integral cross sections.} Experimentally observed ICSs for inelastic ammonia-ammonia (left) or NO-ammonia isotopologues (right) as a function of the collision energy. Theoretically predicted curves are determined from simulations of the experiment taking the electric field-dependent theoretical ICSs as input (see SM). Uncertainty in the simulations is indicated by the colored shaded area. The NO-ND$_3$ data were taken from Ref. \cite{Tang:Science379:1031}. Data is accumulated using a continuous cycle over collision energies. Vertical error bars represent statistical uncertainties at 95\% of the confidence interval (all panels). Horizontal error bars represent uncertainties in collision energy (all panels). Scattering signals were corrected for flux-to-density effects taking the spatial, temporal and velocity distributions of both colliding packets into account (see SM). Experimental cross sections are scaled to the theoretical ones.}	
\label{fig:ICS}
\end{figure}

We measured integral cross sections (ICSs) for inelastic collisions inducing the $|-\rangle \rightarrow |+\rangle$ inversion doublet changing transition for $^{15}$ND$_3 - ^{14}$ND$_3$,  
$^{14}$ND$_3 - ^{14}$NH$_3$ and $^{15}$NH$_3 - ^{14}$NH$_3$, where the first and second species mentioned pertained to the Stark decelerator and curved hexapole beamlines, respectively (see Fig. \ref{fig:ICS}). We detected collision products originating from the Stark decelerator beamline only. For each system, at high energies we observed a cross section that followed the $E_{coll}^{-2/3}$ energy dependence predicted by the Langevin model. Towards lower energies, clear maxima in the cross sections were observed that moved to higher energies for larger values of $\Delta$, which we interpreted as the LM. For ND$_3$ - NH$_3$, the energy window was sufficiently large to cover the low-energy side of the LM, and we observed a plateau forming at energies below $\sim 1$ cm$^{-1}$. For NH$_3$-NH$_3$, the scattering signals dropped to undetectable levels at energies below 4~cm$^{-1}$.  

We converted measured scattering signals into relative cross sections using advanced simulations of the experiment (see SM), and compared the experimental ICSs with theoretical ICSs based on coupled-channels (CC) scattering calculations using an \emph{ab initio} NH$_3$-NH$_3$ potential energy surface (PES) \cite{jing:22}. We limited the channel basis to the $1_1^-$ and $1_1^+$ levels for both molecules, as extension to higher rotational levels was computationally prohibitively expensive. Although convergence with respect to the rotational basis was not necessarily reached, we obtained reliable cross sections as the collision dynamics is dominated by long-range interactions that were still fully captured in our limited basis (see SM). We found good agreement between experiment and theory, validating the existence of the LM and its dependence on $\Delta$. We confirmed that the LM and underlying mechanism is ubiquitous and not specific to ammonia-ammonia collisions by revisiting the NO-ammonia system. The rotational ground state of the NO radical has a parity doublet structure similar to ND$_3$, and for ease of discussion we will also refer to the upper and lower parity component of NO as $|-\rangle$ and $|+\rangle$, respectively.  Unlike our previously measured ICS for NO-ND$_3$ where the LM was located outside the experimentally accessible region \cite{Tang:Science379:1031}, we observed the LM for NO-NH$_3$ and found excellent agreement with the theoretical prediction for this system, see Fig \ref{fig:ICS}.  

To corroborate these findings, we performed model calculations to elucidate the scattering mechanisms across all collision energies probed. We distinguished between two processes that contributed to the total cross section: one where the undetected collision partner changed its parity ($|--\rangle \rightarrow |++\rangle$ collisions), and one where it did not  ($|--\rangle \rightarrow |+-\rangle$ collisions).
We found that for energies above and around the LM, the ICS mainly depended on the dipole-dipole contribution to the interaction potential by coupling the $|-\rangle$ and the $|+\rangle $ states in each molecule, leading to de-excitation collisions in which both molecules flipped their parity (see SM).
As the energy lowered, the classical turning point during the collision shifted to larger intermolecular distances $R$, where the dipole-dipole interaction evolved from a $1/R^3$ to a $1/R^6$ dependence \cite{Tang:Science379:1031}.
Consequently, the mutual polarization resulting from mixing of the $|-\rangle$ and $|+\rangle$ parity states in both colliders could no longer be fully sustained, and the cross section started deviating from the Langevin power law until the LM appeared. In the SM we describe a minimal $2\times2$ model that reproduces CC results almost quantitatively and explains the formation of the LM and the suppression of the cross section at lower energies in terms of suppressed non-adiabatic transitions.

\begin{figure}[!htb]
\centering
\includegraphics[width=0.6\linewidth]{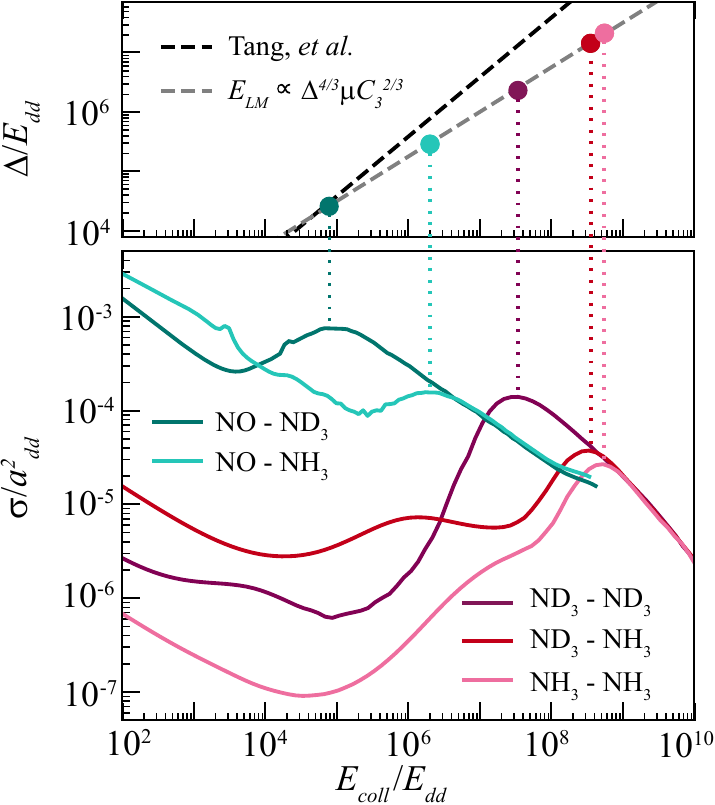}
\caption{{\bf Universality of the LM position.}
Theoretically predicted ICS curves from coupled-channels scattering (lower panel) are cast into dimensionless form using the dipolar length $a_\mathrm{dd}=2\mu C_3/\hbar^2$ and energy $E_\mathrm{dd}=\hbar^2/(2\mu a_\mathrm{dd}^2)$. Peak positions of the LM (upper panel) exhibit universality through the power law $E_\mathrm{peak}/E_\mathrm{dd}\propto (\Delta/E_\mathrm{dd})^{4/3}$. This revisits the earlier findings from Tang \textit{et al.} \cite{Tang:Science379:1031}, where a linear dependence was derived based on a classical Langevin model.}
\label{fig:universality}
\end{figure}

In previous work, the position of the LM was proposed to depend linearly, and solely, on the total splitting
$\Delta=\Delta_A+\Delta_B$ \cite{Tang:Science379:1031}. This scaling law was derived from a Langevin capture model that took
into account the maximum impact parameter $b_\mathrm{max}$ for which collisions at a given energy could still overcome the centrifugal barrier. For large $b_\mathrm{max}$  the dipolar interaction becomes weaker than $\Delta$ and switches off. The LM was taken to occur at the collision energy where the dipolar interaction at capture impact parameter equals $\Delta$. Although this description worked well for NO-ND$_3$, it
was already noted that for systems involving ammonia the model was off by up to an order
of magnitude. This motivated us to revisit the scaling law
by analyzing the results of our coupled channel calculations in dipolar length $a_\mathrm{dd}=2\mu C_3/\hbar^2$, with reduced mass $\mu$ and dipole-dipole coefficient $C_3\sim d_Ad_B$ (see SM), together with the dipolar energy $E_\mathrm{dd}=\hbar^2/(2\mu a_\mathrm{dd}^2)$ to reveal the underlying universality. The resulting theoretically predicted ICSs
for each system are shown in Fig.~\ref{fig:universality}, along with the LM positions as a function of $\Delta$. We found
that the relationship between the peak position $E_\mathrm{peak}$ and $\Delta$ follows the power law $E_\mathrm{peak}/E_\mathrm{dd} \propto
(\Delta/E_\mathrm{dd})^{4/3}$. This revealed that the peak not only depended non-linearly on $\Delta$, but also depended on the masses and dipole moments of each molecule, scaling as $E_\mathrm{peak} \propto \Delta^{4/3}\mu C_3^{2/3}$.

On the low-energy side of the LM, we found that the influence of the dipole-dipole interaction was strongly suppressed, occasionally in favor of the dipole-quadrupole interaction that gained importance. This interaction couples the $|-\rangle$ and $|+\rangle$ states within a single molecule, enabling collisions in which the undetected collision partner scatters elastically ($|--\rangle \rightarrow |+-\rangle$ collisions).
Within our experimentally accessible energy window, this process was particularly strong for ND$_3$-NH$_3$. It exceeded the dipole-dipole contribution by orders of magnitude and even caused a secondary LM to appear (see also Fig.~\ref{fig:universality}).
The experimentally observed plateau for ND$_3$-NH$_3$ collisions at energies well below the primary LM coincided with the onset of this secondary LM, and was attributed to dipole-quadrupole coupling. In general, whether the dipole-quadrupole interaction dominates at low energies is strongly system specific. For example, we identified different universality in scaling of the peak position with $\Delta$ for this secondary LM when colliding molecules with equal or unequal parity splitting (see SM). Interestingly, for NO-NH$_3$ collisions we found that the dipole-dipole and dipole-quadrupole contributions were about equal at a collision energy of $\sim0.1$~cm$^{-1}$.\\

\begin{figure}[!htb]
	\centering
	\includegraphics[width=\linewidth]{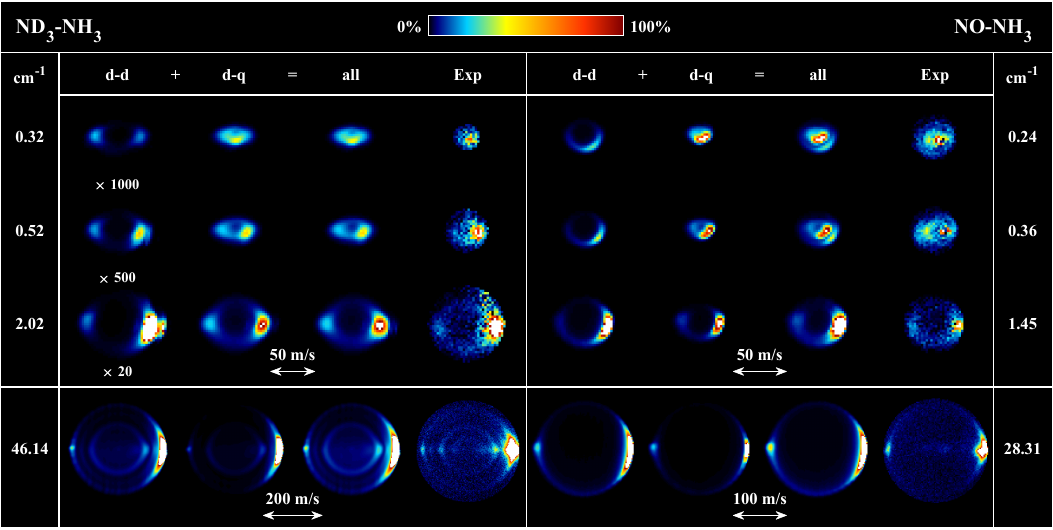}
	\caption{{\bf Collision energy dependence of velocity mapped ion images.} Experimentally observed scattering images for inelastic ND$_3$-NH$_3$ (left) and NO-NH$_3$ collisions (right) as a function of the collision energy, probing the $|-\rangle \rightarrow |+\rangle$ collision-induced transition in the ND$_3$ or NO molecule. Simulated scattering images based on all detectable collisions (all), or taking exclusively the $|--\rangle \rightarrow |++\rangle$ or the $|--\rangle \rightarrow |+-\rangle$ channels into account, which are governed by the dipole-dipole (d-d) or dipole-quadrupole (d-q) coupling, respectively. The intensities of the d-d images for ND$_3$-NH$_3$ were scaled for better visibility. The images are presented such that the relative velocity vector is oriented horizontally, with the forward direction on the right side of the image.}	
\label{fig:DCS}
\end{figure}

As we only detected the collision-induced $|-\rangle \rightarrow |+\rangle$ transition in the molecule from the Stark decelerator beamline, our ICS measurements were blind to the final state of the collision partner.  The ICSs therefore could not discriminate between the $|--\rangle \rightarrow |++\rangle$ and $|--\rangle \rightarrow |+-\rangle$ collision channels, and did not directly probe the individual dipole-dipole or dipole-quadrupole contribution to the interaction. We therefore made complimentary DCS measurements for ND$_3$-NH$_3$ and NO-NH$_3$ using VMI, across collision energies that covered both the low and high energy sides of the LM, see Fig. \ref{fig:DCS}. The idea was that for the $|--\rangle \rightarrow |++\rangle$ process, the de-excitation in the non-detected collision partner would result in a recoil energy that is imparted to the detected molecule. This will increase the radius of the scattering image, in particular at collision energies below or equal to $\Delta_B=$ 0.8~cm$^{-1}$ of NH$_3$, which can be resolved using VMI. By contrast, if the $|--\rangle \rightarrow |+-\rangle$ process dominates, the collision partner scatters elastically, and only the much smaller splitting energy $\Delta_A$ of the detected ND$_3$ or NO molecule is released, which is too small to affect the image sizes even at the lowest collision energies probed.  

At energies above $\sim 20$~cm$^{-1}$, we observed multiple concentric rings, pertaining to the different endo-energetic rotational inelastic excitation channels in ND$_3$ or NH$_3$. A narrow backscattered feature was observed in all images, which we attributed to soft collision backward glories as found before in NO-ND$_3$ collisions \cite{Tang:Science379:1031}. As the energy was reduced, the images became smaller, and converged to a small structureless dot for ND$_3$-NH$_3$ at the lowest energy probed. By contrast, at a collision energy of 0.2~cm$^{-1}$, the image for NO-NH$_3$ showed a central dot with surrounding halo. 

We simulated the expected scattering images based on the predicted cross sections from the full PESs, and found excellent agreement with the experimental images. As the channel basis in our ammonia-ammonia calculations was truncated at $j=1$, we took for the highest collision energy an isotropic DCS to simulate the angular distributions of the inner rings. We found that indeed both the angular distributions and the sizes of the images critically depended on the individual contributions of the dipole-dipole and dipole-quadrupole interactions to the cross sections, in particular at energies below $\sim 1$ cm$^{-1}$. To visualize this, we performed model simulations of the images that would be expected if only the $|--\rangle \rightarrow |++\rangle$ (governed by the dipole-dipole interaction, labeled d-d) or the $|--\rangle \rightarrow |+-\rangle$ (dipole-quadrupole, labeled d-q) collision channels would contribute, artificially normalizing the strength of both channels for clarity (see Fig. \ref{fig:DCS}). 

The measured dot for ND$_3$-NH$_3$ at the lowest energies directly implied the absence of significant recoil energy, and validated the theoretical prediction that NH$_3$ primarily scattered elastically at this energy, consistent with a dipole-quadrupole dominated interaction. By contrast, the dot with surrounding halo measured for NO-NH$_3$ at $E_{coll}=0.2$ cm$^{-1}$ was testament of the bimodal dipole-dipole and dipole-quadrupole nature of the interaction once the energy dropped below the LM for this system. The result that the dipole-dipole interaction becomes less important at low energies for the system with the largest dipole moments is counterintuitive, but is the direct result of the quantum nature of the dipole moment that can significantly suppress the effective dipole moments during a collision.\\

\section*{Conclusion}

The experimental validation of the LM, its universalness for systems with parity doublets, and the appreciation that the dipole-dipole interaction can effectively switch off at low energies even for systems with relatively large dipole moments, may be a double-edged sword for future studies that aim to investigate collisions between cold polar molecules. Certainly, the reduction of inelastic cross sections at energies below the LM may imply bad news for the feasibility of detailed scattering studies in crossed or merged beam scattering experiments. In addition, trap experiments typically operate in the energy regime of the LM, and collision dynamics inside the trap may be significantly affected by its occurrence. Yet, at the same time it offers distinctive opportunities for manipulation and control. The LM as observed here is expected to respond extremely sensitively to external electric fields. For molecules like OH and ND$_3$, the near-degenerate states with opposite parity become already mixed in electric field strengths of 1~kV/cm or less, suggesting that low external fields can drastically modify the cross sections, even for the collision energies probed here that are far above the ultracold limit. Time varying fields can switch on or off the effective dipole moments of the molecules, and could be used to toggle between a dipole-quadrupole or dipole-dipole dominated interaction. The ability to control both the strength and direction of the field, as well as the different angular momentum projection states with quantum number $|m|$ that respond differently to electric fields, will yield additional possibilities for control. Last but not least, if the molecule also possesses a magnetic dipole moment, such as OH or NO in their $X^2\Pi_{3/2}$ state, the parity splitting $\Delta$ can be tuned using magnetic fields. All these possibilities offer unique prospects to tune both the nature and strength of the interaction while probing state-to-state integral and differential cross sections, at collision energies that are now available in advanced crossed and merged beam scattering experiments.    

\section*{Methods}
The experimental setup ran at a repetition rate of 10~Hz. The primary (Stark decelerator axis) beam contained either NO, ND$_3$ or NH$_3$, whereas the secondary (curved hexapole arm) contained either ND$_3$ or NH$_3$. For the ND$_3$-ND$_3$ and NH$_3$-NH$_3$ systems, the Stark decelerator beam axis contained ammonia molecules with isotopically enriched ($> 99\%$ purity) $^{15}$N atoms. 

The primary molecular beam was generated by expanding 5\% ammonia or NO seeded in rare gases through a Nijmegen Pulsed Valve (NPV) at a typical backing pressure of 1 bar. Using different carrier gases from Xe/Kr to Ar/Ne, the Stark decelerator (316 stages, operated at $\pm$17~kV), produced packets of molecules with a tunable mean velocity between 390 m/s and 980 m/s with a narrow velocity spread. After exiting the Stark decelerator, the molecules were guided by a 20~cm long straight quadrupole (operated at $\pm$ 2 kV) until it entered the straight arm of the merged guide (20~cm long). The straight quadrupole was needed to accommodate sufficient space between the Stark decelerator and the curved hexapole (operated at $\pm$ 12 kV), which was used to manipulate the secondary beam. The secondary beam with a mean velocity of $\sim$ 385 m/s was produced by expanding 5\% ammonia seeded in Kr/Xe by a second NPV at a backing pressure of 1 bar. After passing through a skimmer, it was bent by a curved hexapole, and loaded into the curved arm of the merged guide. Upon exiting the guide, the two beams traversed a 1~cm gap and then entered the near-fieldfree region between two electrodes of the ion optics (with radius of 5.5~cm). Based on the merged guide design, the effective crossing angle between the two beams amounted to $2^{\circ}$, affording a collision energy range of 0.2 – 130 cm$^{-1}$. 

The scattered products were state-selectively ionized at the center of the ion optics and detected by VMI. Measurements involving NO used a (1+1$^{\prime}$) REMPI scheme as described before \cite{Tang:Science379:1031}. ICS measurements for ammonia-ammonia employed a convenient one-color (2+1) REMPI scheme, using the 317~nm output ($\sim$ 13 mJ) of a dye laser. For DCS measurements, a more elaborate recoil-free VUV-vis (1+1$^{\prime}$) REMPI scheme \cite{Kuijpers:JPCA128:10993} was needed. Here, VUV light at $\sim$ 160~nm was generated by difference frequency four wave mixing in a cell containing pure Krypton, and used to resonantly excite a rovibrational transition in the $\tilde{B}\leftarrow X$ band of ammonia. The molecules were then further excited to a Rydberg state by a visible photon around $\sim$ 448 nm, and finally autoionized in the electric field of the VMI detector. 

In all measurements, the ionized products were subsequently collected using a high resolution VMI detector, consisting of 16~stacked cylindrical electrodes and a flight tube of 1453 mm length \cite{Plomp:MP119:e1814437}. The design of the ion optics allowed for a low electric field of 20.6 V/cm in the ionization region to limit background signal from undesired parity mixing of the two doublet components \cite{Tang:Science379:1031}. To ensure complete collection of the Newton sphere, and to suppress background contributions from other masses, a pulsed voltage of 300 ns duration was applied to the second microchannel plate of the VMI detector. 

During ICS measurements, the collision energy was scanned using fully automated cycles, in which collision signals and background signals were recorded while the velocity of the Stark decelerated beam was adjusted every few shots of the experiment. For DCS measurements, the Stark decelerator was programmed to produce a certain velocity without making automated scans over the collision energy. The experimentally obtained normalized
signals were corrected for flux to density effects using extensive Monte Carlo simulations
of the experiment (see SI). 

State-to-state cross sections were computed by coupled-channels scattering using the renormalized Numerov method \cite{johnson:78} on \textit{ab initio} CCSD(T) level potentials for NH$_3$-NH$_3$ \cite{jing:22} and NO-NH$_3$ \cite{Tang:Science379:1031}, applying $S$-matrix boundary conditions at large intermolecular distances \cite{janssen:13}. Universality of the LM position was then assessed by expressing the ICS in dipole-dipole units. To interpret the formation of the LM, the coupled-channels equations were reduced to minimal $2\times2$ models, connecting the classical and quantum capture regimes. The secondary LM arising from dipole-quadrupole-driven transitions was analyzed using a distorted-wave Born approximation model (see SI).

\bibliography{references}

\begin{thebibliography}{10}
\bibitem{Note:5}
\bibinfo{note}{In quantum mechanics, a dipole moment arises by the dipole
  operator that has odd parity. The expectation value of the dipole momemt for
  a molecule in a given quantum state with even or odd parity is always zero
  (i.e., $\langle +|d|+ \rangle = \langle -|d|- \rangle = 0$).}
\bibitem{Note:4}
\bibinfo{note}{We use the wavenumber (cm$^{-1}$) unit for collision energy
  throughout, as molecules in molecular beams are not in thermal equilibrium.
  One wavenumber corresponds to a temperature of 1.4 Kelvin.}
\end{thebibliography}
\bibliographystyle{naturemag}

{\bf Acknowledgement}
We thank Youp Caris for help during the initial phases of the project, and Niek Janssen for expert technical support.
{\bf Funding:}  This work is part of the research program of the Netherlands Organization for Scientific Research (NWO) through an ENW-M and a VICI grant. S.Y.T.v.d.M. acknowledges support from the European Research Council (ERC) under the Horizon Europe's program (Grant Agreement No. 101141163 — QUCUMBER). {\bf Author contributions:} The experiments were performed by Y.C. and A.v.R. with initial contributions by S.H., and analyzed by Y.C. and S.K.. The theoretical calculations were performed by E.W. and T.K.. Simulations on beam overlap were performed by Y.C. and S.K. The project was conceived and supervised by S.Y.T.v.d.M.. The main manuscript was written by S.Y.T.v.d.M.. The Supplement was written by Y.C., E.W. and S.Y.T.v.d.M.. All authors were involved in the interpretation of the data, discussed the results, and commented on the manuscript. {\bf Competing interests:} None declared. {\bf Data and materials availability:} All  data needed to evaluate the conclusions in the paper are present in the paper or the Supplementary Materials, and will be made available online at DOI: 10.5281/zenodo.21620550.
\newpage

\end{document}